\documentstyle[twocolumn,aps,epsfig] {revtex}
\begin{document}
\bibliographystyle{unsrt}

\title{Classical Drop Phase Diagram and Cluster Distributions}
\author{A. Chernomoretz, P. Balenzuela, C.O. Dorso}
\address{Departamento de Fisica, Universidad de Buenos Aires,
Buenos Aires, Argentina.}

\date{\today}
\maketitle
\begin{abstract}
The phase diagram, ($T,\rho$), of a finite, constrained, and classical system is built
from the analysis of cluster distributions in phase and configurational
space. The obtained phase diagram  can be split in three regions. One,
low density limit, in which first order phase transition features can be
observed. Another one, corresponding to the high density regime, in which
fragments in phase space display critical behavior of 3D-Ising
universality class type. And an intermediate density region, in which
power-laws are displayed but can not be associated to the abovementioned
universality class. 
\end{abstract}
PACS number(s): 25.70 -z, 25.70.Mn, 25.70.Pq, 02.70.Ns

\section{Introduction}

The multifragmentation phenomenon that takes place in nuclei with excitation
energies above 2 MeV/nucleon has been one of the central issues of the
nuclear
community during the past two decades. Many experimental and theoretical
efforts have been concentrated towards the understanding of the mechanisms
involved in that process. In particular, one feature that triggered the
interest on the field was the important detected production of intermediate
mass fragments (IMF's). Highly excited nuclei break up in many IMF's
in Fermi energy range reactions. This feature, along with the fact that early
calculations of caloric curves showed an approximately
constant behavior, was interpreted by many groups as
a signature of a phase transition taking place in finite nuclear systems
\cite{nuovo,pocho95,gross,gulminelliDurand}.
 In recent contributions the behavior of the microcanonical
heat capacity was used to experimentally characterize the transition as a
first order
one \cite{dagostino}. Within this picture, the latent heat can be associated
with the transformation between a Fermi liquid and a gas phase, composed by 
light particles and free nucleons.

Despite that the familiar liquid-gas transition framework seems to be
appropriate to
deal with the nuclear case, there are some peculiarities that are worth to
be considered. One key point usually disregarded is that, by its own nature,
multifragmentation in nuclear reactions should be {\em a priori} analyzed as
a nonequilibrium process (see \cite{Strachan97,cherno}).  Even if an
equilibrium scenario is adopted, a major point is that we are dealing with a
phase transition occurring in a finite system. Therefore, several
thermodynamical features, e. g.  the entropy extensivity,
cannot be taken for granted and the agreement between
different statistical ensembles predictions cannot be invoked
anymore \cite{grossbook}.
A working hypothesis, usually made by statistical models in order to apply a
tractable global equilibrium picture, is that a freeze-out volume can be
defined inside of which the existence of an equilibrated ensemble of
clusters can be assumed. In these models, the behavior of thermodynamics
quantities is closely related to the way the system is {\em partitionated} into
clusters giving rise to internal surfaces (isolated drops).

The aim of the present paper is to analyze the relationship between cluster
distributions inside a fixed volume and the respective thermodynamic 
description. In order to do that a
confined system that interacts via a two-body Lennard-Jones
potential is studied. A molecular dynamics, MD, approach is used.
The analogy between the nuclear force and the van der Waals
interaction supports the use of this simplified classical model to
obtain qualitatively meaningful results.

Given the MD microcanonical description, a connection
with thermodynamical quantities can be established via well known
 principles from mechanics. For instance, the pressure and temperature
can be estimated from the generalized equipartition theorem \cite{cagin},
whereas the specific heat can be related to kinetic energy
fluctuations \cite{percus}.  On the other hand, a complementary analysis
of the microscopic correlations at a ``nucleon'' level of description can
also be considered. To that end, different fragment-recognition
algorithms can be used in order to unveil different particle
correlation properties.

Previous studies \cite{campi1y2} have already dealt with
this 'cluster structure-thermodynamic description' mapping. They made used of
a cluster definition \`a la Coniglio-Klein, and were mainly focused
on the system behavior in the supercritical phase ($\rho>\rho_c,T>T_c$).

In this contribution we consider two alternative fragment
definitions that, differently from Coniglio-Klein clusters (that 
in the present paper will be
called MSTE clusters, as will be explained in Sec. \ref{secFrag}),
are built in
well defined and physically meaningful spaces.
The first one, associated with the so called minimum spanning tree fragment
recognition method, is based on configurational information (MST
clusters)\cite{mst}. The second one uses complete phase space information in order to
define a fragment set according to the most bound density fluctuation in phase space
\cite{dorso93} (ECRA clusters).
Using this fragment characterization as a fundamental piece
of information, consistent phase diagrams are built from scratch
for the analyzed finite systems. In addition, the relationship between
critical signatures in the fragment distributions and the corresponding system
localization in the phase diagram is established.

This paper is organized as follows. In Section~\ref{secModel} we will
describe the model used in our simulations. Section~\ref{secFrag} is devoted to
a characterization of the used cluster definitions.
A description of expected signals associated with phase transitions in
finite systems is given in Section \ref{secSig}. In Section~\ref{secPD} the study of
phase diagram is presented. Finally, in Section~\ref{secConc}, conclusions are drawn.

\section{The model}\label{secModel}
\smallskip The system under study is composed by excited drops made
up of particles interacting via a 6-12 Lennard Jones potential, which
reads:

\begin{equation}
V(r)=\left\{
\begin{array}{ll}
4\epsilon \left[ \left( \frac{\sigma }{r}\right) ^{12}-\left(
\frac{\sigma }{
r}\right) ^{6}-\left( \frac{\sigma }{r_{c}}\right) ^{12}+\left(
\frac{\sigma
}{r_{c}}\right) ^{6}\right] r \le r_{c} &  \\
0 r$>$r_{c} &
\end{array}
\right.
\end{equation}

\smallskip We took the cut-off radius as $r_{c}=3\sigma $. Energies
and distances are measured in units of the potential well ($\epsilon $)
and the distance at which the potential changes sign ($\sigma $),
respectively while the unit of time used is:
$t_{0}=\sqrt{\sigma ^{2}m/48\epsilon }$.

In order to constrain the dynamics we used a spherical confining `wall'.
The considered external potential behaves like
$V_{wall} \sim (r-r_{wall})^{-12}$ with a cut off
distance $r_{cut}=1\sigma $ , where it smoothly became zero along with
its first derivative.
The set of classical equations of motion were integrated using the well
known velocity Verlet algorithm \cite{Verlet}, taking $t_{int}=0.002t_{0}$
as the integration time step. Once the transient behavior was over  a
microcanonical sampling of particle configurations every $5t_{0}$ up
to a final time of $140000t_{0}$ was performed.

\section{Fragment Definitions}\label{secFrag}

The simplest and more intuitive cluster definition is based on
correlations in configuration space: a particle $i$ belongs to a
cluster $C$ if there is another particle $j$ that belongs to $C$
and $|{\bf r_i}-{\bf r_j}| \leq r_{cl}$, where $r_{cl}$ is a
parameter called the clusterization radius. 

We set $r_{cl}=r_{c}=3\sigma$. 
The algorithm that recognizes these
clusters is known as ``Minimum Spanning Tree'' (MST).
In this method only correlations in {\bf q}-space are used,
neglecting completely the effect of momentum.
As was shown in Ref.~\cite{mst}, MST clusters give incorrect information about 
the meaningful fragment structure of the system for dense
configurations. 
However, an interesting point to be notice is that it
 can still provide useful information about the limit imposed
by the constraining finite volume to the formation of well defined
fragments in configurational space. As $r_{cl}=r_{cut}$, no inter-cluster
interaction exists, so cluster surfaces can be univocally defined.

An extension of the MST is the ``Minimum Spanning Tree in Energy''
space (MSTE) algorithm. In this case, a given set of particles
$i, j,..., k$, belongs to the same cluster $C_i$ if:
\begin{equation}\forall \, i \, \epsilon \, C_i \:,\: \exists \, j \,
\epsilon \, C_i  \, /
\,  e_{ij} \leq 0
\end{equation}
where $e_{ij} = V(r_{ij}) + ({\bf p}_i - {\bf p}_j)^2 / 4 \mu$,
and $\mu$ is the reduced mass of the pair $\{i,j\}$. This 
cluster definition resembles the clusterization prescription adopted by
Coniglio and Klein \cite{Coniglio} and was the one used by Campi {\em et al.}
in Ref.\cite{campi1y2}.
The MSTE algorithm
searches for configurational correlations between particles
considering the relative momenta of particle pairs,
and typically recognizes fragments earlier than
MST in highly excited unconstrained systems\cite{mst,lj2d}.

Finally, a more robust cluster definition is based on the system ``most bound density
fluctuation''(MBDF) in phase space \cite{dorso93}.  The MBDF is composed by
the set of clusters $ \{ C_i \}$ for which the sum, $E_{ \{C_i\}}$, of the fragment internal
energies attains its minimum value:
\begin{eqnarray}
E_{ \{C_i\}} &=& \sum_i E_{int}^{C_i}
\nonumber \\
\mbox{with    } E_{int}^{C_i}& = & \sum_{j \in C_i} K_j^{cm} + \sum_{ {j,k
\in C_i} \atop j \le k} V_{j,k} \label{eq:eECRA}
\end{eqnarray}

 $K_j^{cm}$ is the kinetic energy of particle $j$
measured in the center of mass frame of the cluster which contains
particle $j$, and $V_{ij}$ stands for the inter-particle
potential.

The algorithm that finds the MBDF is known as the ``Early Cluster
Recognition Algorithm'' (ECRA).  
It searches for simultaneously well correlated structures in both, {\bf q}, 
and {\bf p} space, via the minimization of the potential and the kinetic terms 
respectively.

The ECRA algorithm  has been used extensively in many studies of free expanding 
fragmenting systems ~\cite{Strachan97,ecras} and has helped to discover that
excited drops break very early in the evolution.  In addition, in a recent
contribution \cite{cherno} it was shown that ECRA clusters are also suitable to
describe the fragmentation transition that takes place in volume constrained systems.

\section{Characterizing the transition}\label{secSig}

As mentioned in the introduction, several observables can
be studied in order to analyze phase transitions phenomena
occurring in finite systems. They usually involve the behavior of caloric
curves
\cite{pocho95,gross,pochodzalla,lynden,wales,richert}, specific heats
\cite{dagostino,nacluster}, kinetic energy fluctuations
\cite{gulminelliDurand,chomaz}, fragment
mass distributions \cite{purdue,elliott}, and critical exponents
\cite{elliott}.

\subsection{Thermodynamical Description}\label{termo}
One of the most relevant quantities in the study of fragmenting systems,
either constrained or free to expand is the caloric curve (CC).
The CC is defined as the functional relationship between the
system energy and its temperature, given by:
\begin{equation}
T(E) = {2 \over {3 (N-1)}} <K>_E \label{eqAclara}
\end{equation}
being N the number of particles, and $<K>_E$ the mean kinetic energy averaged
over a fixed total energy MD simulation \cite{aclara}.

\begin{figure}
\centerline{\epsfig{figure=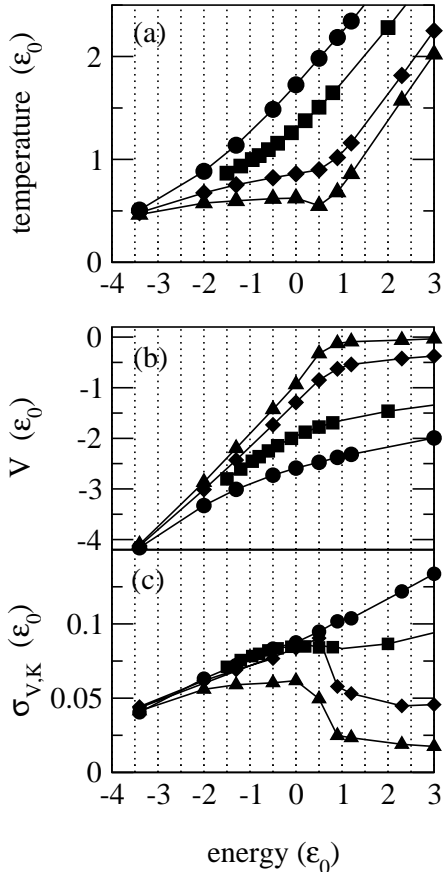,height=12cm}}
\caption{The caloric curve is shown in panel (a). In panels (b) and (c) the potential
 energy ($V$) and its root mean squared deviation as a function of the system 
total energy are displayed in panels (b) and (c) respectively. The following density
values: $\rho = 0.01 , 0.07, 0.35,$ and $0.55 \sigma^{-3}$, displayed with full
triangles, diamonds, squares, and circles respectively, were considered.}\label{figCC}
\end{figure}

In Figure~\ref{figCC}(a) the CC's are shown for the following densities: 
$\rho^h=0.55 \sigma^{-3}, \rho^c=0.35 \sigma^{-3}, \rho^m=0.07 \sigma^{-3},$ and $\rho^l=0.01 \sigma^{-3}$.

Different behaviors can be recognized.
For the more diluted case, $\rho^l$ (full triangles), the corresponding
CC displays a loop which ends in a linearly increasing temperature line,
which we refer as {\em vapor branch}.
Taking into account that, within the microcanonical ensemble, the specific 
heat is defined as:
\begin{equation}
{1\over{C}} = \frac{\partial T}{\partial E} = -T^2 \frac{\partial^2 S}{\partial
E^2} \label{eqC}
\end{equation}

it is clear that for this case negative values of $C$ will be found in the range 
$0.1 \epsilon_0 < E < 0.6 \epsilon_0$.
A negative value of the derivative of the temperature as a
function of the energy reflects an 'anomalous' behavior of the system entropy
for the respective energy range.
A {\em convex intruder} in $S(E)$, prohibited in the thermodynamical
limit, arises as a consequence of the finiteness of the system and  the
corresponding lack of extensivity of thermodynamical quantities like
$S$. This signal is expected in first order phase transitions, and is
associated to a negative branch of the heat capacity between
two poles \cite{grossbook}.
Taking into account the relationship between kinetic energy fluctuations and
the system specific heat \cite{chomaz,lebowitz}:
\begin{equation}
{1 \over N} <(\delta K)^2>_E={3 \over {2 \beta^2}} (1 - {3 \over {2 C}}) 
\label{eqFluc}
\end{equation}
it can be seen that negative values of the specific heat appear whenever
$(\delta K)^2$ get larger than the canonical expectation:
${3 N\over{2 \beta^2}}$.
This behavior has already been verified in Ref.\cite{cherno}.

As the density is increased the loop is washed away and it is replaced
by an inflection point ($\rho^m$, full diamonds in Fig.\ref{figCC}(a)).
This corresponds to the merging of the specific heat negative poles into 
a single local maximum.
Finally, for higher densities, $\rho^c$ and $\rho^h$ 
(filled diamonds, and circles), no major changes in the CC's second derivative 
is observed. 

In Fig.~\ref{figCC}(b) the mean potential energy per particle, $V$, as
a function of the total system energy is displayed. 
For the lowest densities two different regimes can be recognized.
First, a steep increasing behavior for energies $E<0.5\epsilon_0$, that 
can be related to an increase of the mean interparticle distance and the 
appearance of internal surfaces, as the number of atractive bonds decreases.
Second, a saturating
behavior for larger energy values, $E>0.5\epsilon_0$. 
For more dense systems, no such feature can be observed. A smooth
behavior is displayed by $V$.

Another interesting feature can be noticed looking at the root
mean square deviation of partial energies (kinetic or
potential), $\sigma_{V,K}$,
shown in Fig.~\ref{figCC}(c). For the lowest densities (filled triangles
and diamonds) an abrupt decrease is observed at $E\sim0.5\epsilon_0$,
whereas for the highest considered density,
$\rho^h=0.8\sigma^{-3}$(filled circles), no trace of such behavior can
be reported. For future reference, it is worth noting that the
transition in the monotonic character of the curve takes place at 
a density value $\rho\sim\rho^c=0.35\epsilon_0$(filled squares).

To attain a better understanding of the behavior of the magnitudes displayed 
in Figs.~\ref{figCC}(b),
and (c), we have followed two strategies: i) to study particle-particle
correlations, ii) to analyze fragment mass distributions according to the
already presented fragment definitions. In what follows we present
results obtained within the first approach, and postpone to the next
subsection the analysis of fragment properties.

\begin{figure}
\centerline{\epsfig{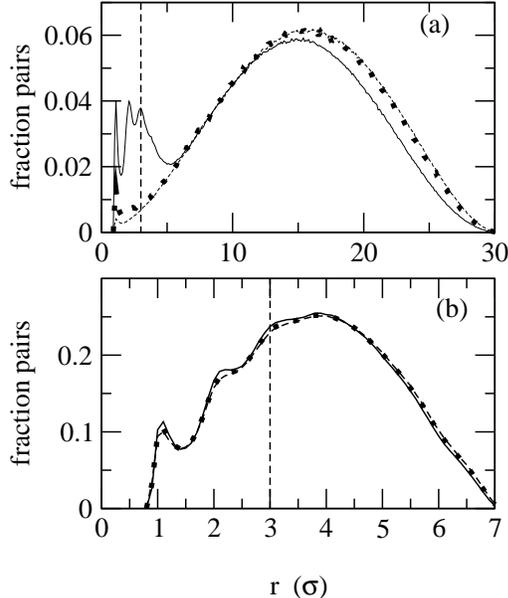}}
\caption{Normalized interparticle distance distribution calculated for
$\rho^l$ and $\rho^h$ systems are shown in panels (a) and (b)
respectively. The following three total energy values, $E=0.0, 0.5, 1.0\epsilon_0$, 
were considered in panel (a), whereas $E=-0.50, 0.5, 1.0\epsilon_0$ for panel (b).
They are shown with, solid, dotted, and dashed lines respectively.
Note the different scales used for $r_{ij}$ in both panels.}\label{figRij}
\end{figure}

In Fig.\ref{figRij} the distribution of interparticle distances,
$r_{ij}$, is shown. Two density values already considered in Fig\ref{figCC},
namely $\rho^h$ and $\rho^l$, are shown in panels (a) and (b) respectively. 

For each density, three total energy values, lower than(solid line), 
equal to (dotted line), and larger than (dashed line) $E=0.5\epsilon_0$, 
were considered (see caption). In both panels, a vertical line was included 
indicating the interaction cut-off radius.

In panel (a) (very diluted case) a well defined structure of interacting particle
pairs can be seen  at low energies (continuous line). 
The displayed peaks signal the presence of a rather large self-sustained drop,
with a first, second and even third nearest neighbors structure. 
As energy increases, this structure fades away, and an important reduction of the number of 
interacting particles can be noticed. This occurs as a consequence of the spacious 
volume available
, that is big enough to accommodate small non-overlapping
particle aggregates. In this transition, well defined {\bf q-space} clusters appear,
surfaces are produced and $V$ tends towards a residual value (see
bellow). 

A completely different behavior is observed for the denser
instance, $\rho^h$, shown in Fig.~\ref{figRij}(b). 
In this case, no changes in the statistical distribution of $r_{ij}$ is observed as 
the energy is increased. (Note that the bell shaped distribution associated with
the `trivial' noninteracting pair counting is superimposed over the
 peak structure. However, the presence of a first, second, and even a third nearest
neighbors can still be traced).
The container imposes a severe volume restriction on the system,
and even for high total energy values, each particle is confined
in the attractive concavity of partners potential, between the repulsive 
core of nearest neighbors.

This last observation can complete a general picture, within which the
behavior of the mean potential energy, $V$, for the high density cases can be easily
understood. As more energy is added to the system no structural transition is
allowed by the constraining volume. 
Particles can not escape from neighbors most attractive potential 
range and a smooth increased in $V$, related to the average time spent in the 
most negative potential areas, can be seen (Fig~\ref{figCC}(b)).

The decreasing of $\sigma_V$, observed  for low densities (filled triangles, and
diamonds in Fig.~\ref{figCC}(c)), can be associated to the loss 
of `attractive bonds' between particle pairs that takes place when a non interacting 
light cluster regime dominates. 
As an increasing number of particles stop
interacting with one another, the system dynamics gets 'less chaotic' (see
\cite{mison} for a dynamical characterization of this system), and a decrease
of $\sigma_V$ is induced.

On the other hand, for high densities, the region with the strongest non-lineal nature
becomes the most relevant part of the interaction potential. As a consequence
of that, an increase of the fluctuations in potential energy between successive
configurations as total energy is added can be expected (see the behavior displayed
by $\sigma_V$ in Fig~\ref{figCC}(c)).

It is worth noting that the presented picture can also be used to
interpret recent results regarding the behavior of the maximum Lyapunov
exponent, MLE, in constrained systems (see \cite{mison}).
Moreover, a striking similitude between the behavior of
$\sigma_V(E,\rho)$ and $MLE(E,\rho)$ can be noticed, comparing
Fig.\ref{figCC}(c), and Fig.5 of Ref.~\cite{mison}.

\subsection{Fragments Inside the Volume}

In section \ref{secFrag}, three fragment recognition algorithms
were described. Each one of them make use of different correlation
information in order to link particles into clusters, and then give
different physical information about the system under study.

\begin{figure}
\centerline{\epsfig{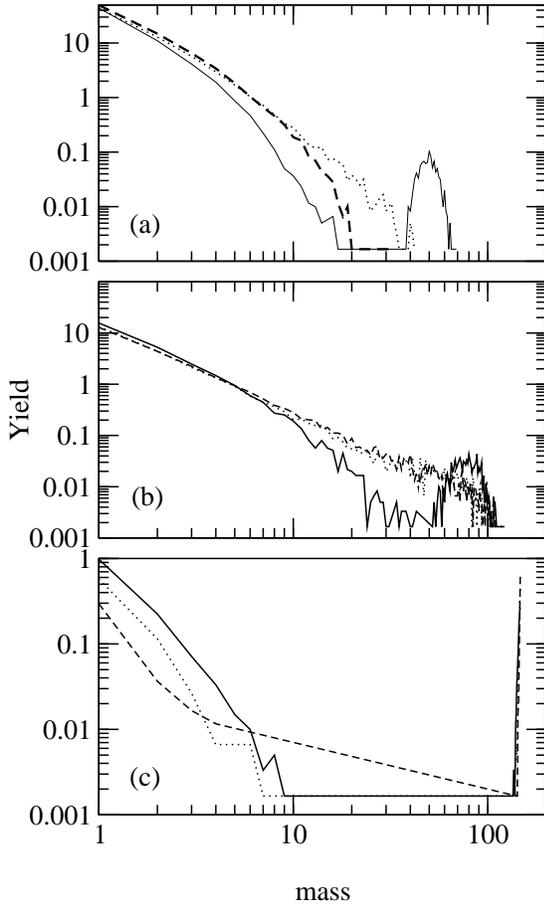}}
\caption{MST cluster mass distributions calculated for densities: $\rho=0.01, 0.03$, and $0.10 \sigma^{-3}$ are displayed in panels (a), (b), and (c) respectively.
Three total energy values were considered for each density,
$E_{tot}=0.0, 0.5$, and $1.0\epsilon_0$.They are displayed as full, dotted and dashed
lines respectively}\label{figClustersMST}
\end{figure}

In Fig.~\ref{figClustersMST} the results of the MST
algorithm analysis are shown. MST spectra
were calculated for different energies ($E=0.0, 0.5$, and $1.0 \epsilon_0$, displayed as
full, dotted and dashed lines respectively) for three system densities:
$\rho=0.01, 0.03$, and $0.10 \sigma^{-3}$. They are shown in panels (a), (b), and
(c), of Fig.\ref{figClustersMST} respectively.

%
 
 At low densities, panel \ref{figClustersMST}(a),
the system evolves from a heavy cluster dominated behavior at low energies, towards 
a light cluster dominated behavior, at high energies. 
This reflects the fact that the constraining volume is large enough to
allow the system to fragment in well defined drops as energy is
increased.
It is interesting to notice that results displayed in
Fig.~\ref{figClustersMST}(a) correspond to full triangle symbols in
Fig.\ref{figCC}(a), i.e. the one for which the CC displays a loop.
It was argued that big fluctuations in kinetic energy should be
expected. This is indeed the case and it is related to the fact that 
well defined surfaces appear in the system (see \cite{cherno}).

On the other hand, as density increases, 
the mass distributions converge to a u-shaped one.
Almost no spatially well separated structures 
(i.e. MST clusters) can be identified, aside from the trivial huge cluster 
that comprises almost all the system particles.

\begin{figure}
\centerline{\epsfig{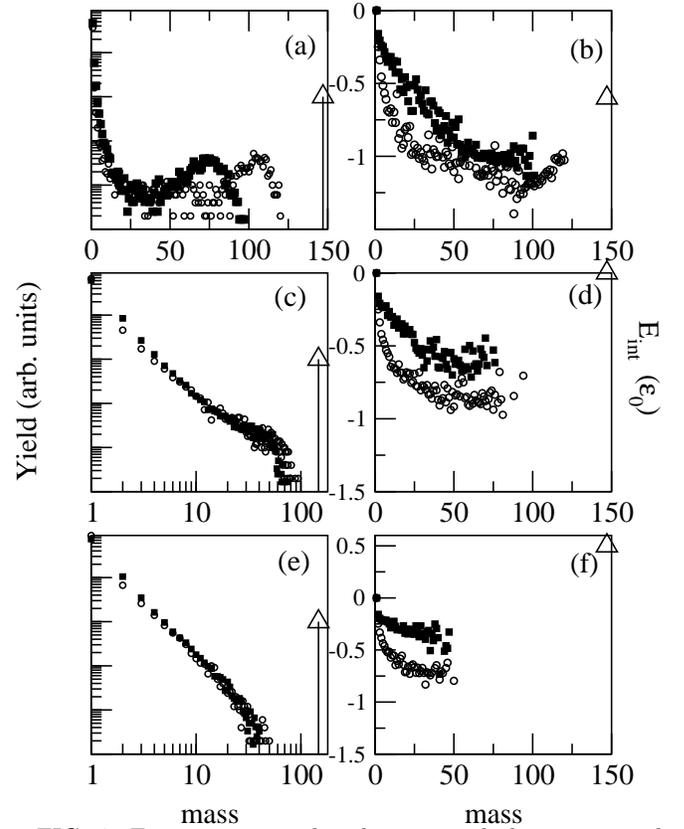}}
\caption{Fragment mass distribution, and cluster internal energy values as a 
function of the cluster mass, are
shown in the first and second columns respectively. Three system total
energies are considered: $E_{tot}=-0.6\epsilon_0$, panels (a-b), $0.0\epsilon_0$, 
panels (c-d), and $0.5\epsilon_0$, panels (e-f). Empty circles, filled
squares, and empty triangles, correspond to ECRA, MSTE, and MST data,
respectively.
}\label{figClusters}
\end{figure}

In Fig.~\ref{figClusters}, the effect of taking into account momentum 
correlation in the definition of clusters is displayed.
There, a cluster analysis for a system of $N=147$ particles
with $\rho=0.2 \sigma^{-3}$ using ECRA, MSTE, and MST prescriptions is
presented.
 The mass distribution function, and the fragment internal energy as a function 
of the cluster mass, $E_{int}(s)$, are shown in the first 
and second columns, respectively. Three total energy values were considered: 
$E_{tot}=-0.6$ (panels (a)-(b)), $0.0$ (panels(c)-(d)), and $0.5 \epsilon_0$ (panels (e)-(f)). Empty triangles, solid squares, and empty circles correspond to MST, MSTE, and ECRA
results respectively.

It can be seen that, for this density, the MST prescription finds just a
big drop. On the other hand, both, the MSTE and ECRA algorithms, 
find non-trivial cluster
distributions that show the expected transition from a u-shaped, towards 
and exponential behavior, as the system energy is increased.
As a general rule, ECRA algorithm produces
 more bound clusters than the MSTE one. This is
consistent with the claim that the MBDF are found by the ECRA algorithm.
Due to this feature, from now on, 
just the properties of the system according to its ECRA cluster structure will be
considered.

\subsection{Transition Signals}

As seen in Fig.\ref{figClusters}, in which a system at a 
$\rho=0.2\sigma^{-3}$ density was considered, the ECRA cluster distributions
undergo a transition from a U-shaped
spectrum towards an exponentially decaying one.
Moreover, the one corresponding to the intermediate energy
value~\ref{figClusters}(b), displays a power law like behavior. 
In fact, for any other density studied, the same behavior can be
reported, and a total energy value can be determined, for which a power-law like mass
distribution can be found. 

This result is interesting, because it implies that some kind of
transition between two regimes, one with, and other without large ECRA drops,
i.e. with or without the 
presence of liquid-like structures, is taking place in high density systems. 
Moreover, this transition can not be detected simply using spatial correlations
information. In this way, considering the appropriate cluster prescription, power-law mass 
distributions can be used to trace the transition line in a
$(T,\rho)$ diagram. 

Within the Fisher Droplet Model \cite{fisher}, when dealing with
infinite systems, one expects to find such a scale-free distribution at a
critical point, where a continuous transition takes place, and scaling
hypothesis can be applied.
 However, it has been reported~\cite{richert,crit1sta,crit1stb}
that several critical signatures also appear when
first order phase transition are analyzed in small systems.
In particular, as stated in Ref~\cite{crit1stb}, in small systems the
largest cluster gets a size comparable to the vapor fraction before dissapearing
when the system crosses the coexistence line. Therefore there is an
energy for which a pseudo-invariance of scale and a resemblance
with critical behavior can be expected for small systems undergoing
phase transitions that, in the thermodynamical limit, would be univocally
classified as first order ones.

In order to search for a power-law behavior in fragment distributions
the following single parameter function was employed
to fit fragment mass spectra. The contribution of the largest
fragment was disregarded~\cite{elliott}:

\begin{equation}
n(A)=q_0A^{-\tau} \mbox{  ,with } q_0(\tau)=\frac{1}{\sum_A
A^{1-\tau}}\label{eqPL}
\end{equation}

$n(A)$ is the number of fragments with mass number $A$, and $q_0$ is a
normalization constant.

\begin{figure}
\centerline{\epsfig{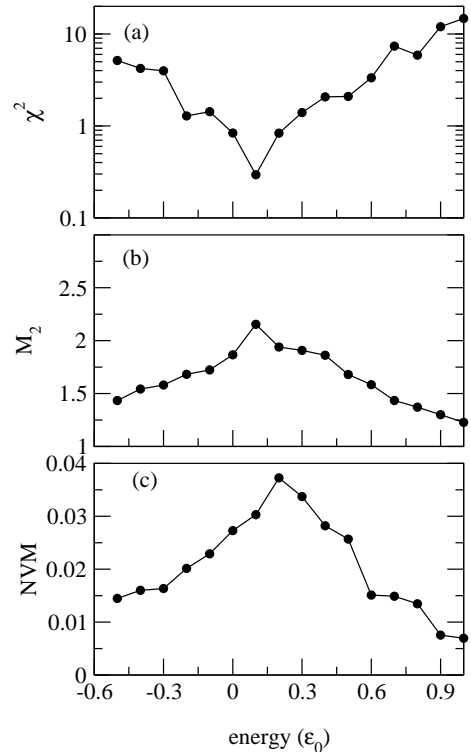}}
\caption{Transition signatures calculated for a $\rho=0.10\sigma^{-3}$
system. The $\chi^2$ fitting coefficient, the second moment of the mass
distribution, $M_2$, and the normalized mean variance of the largest
fragment, NVM, are displayed in panels (a), (b), and (c) respectively.             }\label{figSignals}
\end{figure}

The quality of the fitting procedure was quantified using the standard
$\chi^2$ coefficient (see~\cite{balen} for details), 
and an energy value, $E_{*}$, was associated to the best fitted spectra.
In Fig.\ref{figSignals}(a) a typical $\chi^2$ calculation is shown for
ECRA clusters in a $\rho=0.10\sigma^{-3}$ system. 
From the figure, a value of $E_{*}=0.10
\pm 0.05 \epsilon_0$, can be reported.  

Another useful observable to search for scale-free distribution functions 
is the second moment of the cluster distribution $M_2=\sum'_A A^2 n(A)$.
As in the determination of $\tau$, the largest cluster is excluded from the 
primed sum. $M_2$ is proportional to the compressibility $\kappa_T$, that, in the
thermodynamical limit, presents a power-law singularity at a second
order transition. 
For finite system the divergence is replaced by a maximum.
In panel \ref{figSignals}(b) it is shown the estimation of $M_2(E)$, for
a $\rho=0.10\sigma^{-3}$ system. It can be noticed that the maximum is
located at the energy value for which the spectra is best fitted by a
power-law like dependence.

This kind of agreement is also achieved when the normalized mean
variance, NVM, of the largest fragment mass, $A_{max}$, is analyzed. NVM is defined
as:
\begin{equation}
{\mbox NVM}= \frac{<A_{max}^2>-<A_{max}>^2}{<A_{max}>}
\end{equation}

and it proved to be a robust tool for the characterization of transition
phenomena in which an enhancement of fluctuations occurs\cite{nvm}. As
can be seen from panel \ref{figSignals}(c), this signal, although
slightly shifted within the working resolution, also peaks in
the same region of the previous ones. 

The signal agreement reported for the $\rho=0.10\sigma^{-3}$ case in
Fig.\ref{figSignals}, is
also achieved for every density analyzed in this paper.
This means that, in the whole density range, power-law mass
distributions for phase space defined ECRA clusters are found whenever large 
fluctuations take place in the system.

\begin{figure}
\centerline{\epsfig{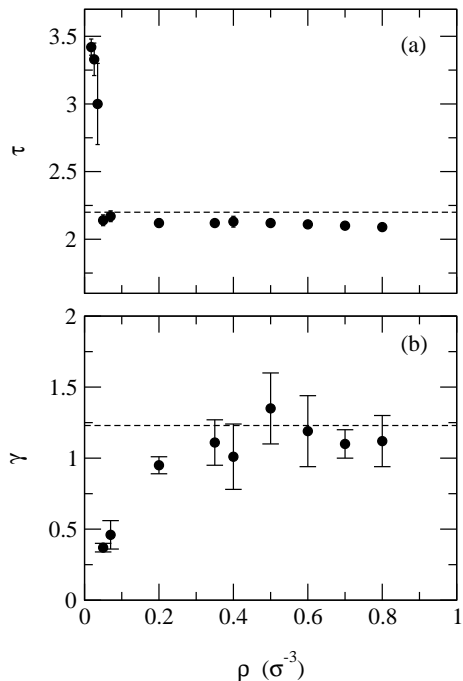}}
\caption{Estimated values from ECRA cluster distributions for the critical 
exponents $\tau$ and $\gamma$ are shown in panels (a), and (b)
respectively. The expected values for the 3D-Ising universality class
are displayed by a dashed line.}\label{figTau}
\end{figure}

In order to properly characterize the state of the system which displays
power-law mass distributions, the values attained by the $\tau$ exponents
must be analyzed, as a function of the density.
This dependence is shown in Fig.\ref{figTau}(a).
For a true critical phenomenon, taking place in three dimensional systems,
$2\le\tau\le3$ is expected \cite{fisher}. This condition is not satisfied for 
 systems with $\rho < 0.05 \sigma^{-3}$. For this highly diluted systems, 
the observed free-scale distribution is not expected to survive the
thermodynamic limit. It is not related to any continuous transition, but
arises as a finite-size effect. 
It is worth noting that $\rho\sim0.05\sigma^{-3}$ is the maximum density
for which the respective caloric curve displays a loop, and then,
negative $c_v$ (see Fig.\ref{figCC}).

On the other hand, for $\rho>0.05$, the calculated $\tau$ exponents show
a rather good agreement with the $\tau=2.21$ value (dashed line), expected for 
liquid-gas transitions.

In a second order phase transition, the behavior of $M_{2}$ near
a critical point can be described in terms of the critical exponent
$\gamma $ \cite{elliott}:

\begin{equation}
M_2(\epsilon) \propto |\epsilon|^{-\gamma}.\label{eqGamma}
\end{equation}    

where, $\epsilon=(Ec-E)/E_c$, measures the distance to the critical
point.
This relation is valid for an infinite system in the limit $\epsilon
\rightarrow 0$. As already mentioned, in a finite system $M_{2}$
displays a maximum instead of the divergence predicted in equation (\ref{eqGamma}).
Having this in mind, a calculation procedure introduced in Ref.\cite{elliott2} 
($\gamma$-{\em matching}), that takes care of finite size effects, was
used to calculate the $\gamma$ exponent value for our system (see also 
Ref.\cite{balen} for details).

The results of the $\gamma$ exponent estimation is shown in
Fig.\ref{figTau}(b).
The obtained values tends toward 
$\gamma_{liq-gas}=1.23$ (dashed line) value, that is expected for a 
liquid-gas transition.
It is worth noting that this convergence is achieved for densities 
$\rho \ge 0.35 \epsilon_0$ (this value equals $\rho^c$ in Fig.\ref{figCC}). 
For this density, a change in the behavior of
the potential energy fluctuations as a function of the total energy was
reported in Fig.\ref{figCC}(c). This can be associated (see Sec.\ref{termo})
to the onset of the invariance of the statistical interparticle distance
distribution (Fig.\ref{figRij}), that takes place as a consequence of the
imposed volume restriction, and precludes the occurrence of a continuous
transition.

\section{Phase Diagram}\label{secPD}

In the previous section several signatures of a change in the properties of
ECRA fragment mass distributions were analyzed.
At any given density, a system energy, $E_*(\rho)$, can be determined 
for which large fluctuations appear in the system, and a scale-free like mass
distribution can be found.

\begin{figure}
\centerline{\epsfig{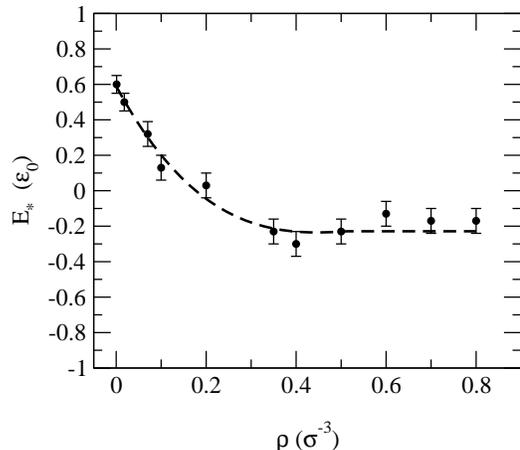}}
\caption{Density dependence of the system energy, $E_{*}$, at which
transition signals are detected. The dashed line is included to
guide the eye.}\label{figEc}
\end{figure}

In Fig.\ref{figEc}, the dependence of $E_*$ with the system density can
be seen. It can be noticed that for $\rho\ge 0.35 \sigma^{-3}$, a rather
constant value is attained for the transition energy. A similar result
was reported in Ref.\cite{campi1y2}, using the already presented MSTE
cluster definition.

Some insight about the properties of the system along the line depicted in
Fig.\ref{figEc}, can be gained from the analysis of the 
interplay between the mean internal potential energy per particle stored 
in ECRA clusters, $V_{int}$, and the mean inter-cluster interaction energy,
$V_{ic}$.

\begin{eqnarray}
V_{int} &=& \sum_{i < j \atop {i,j \in C_k}} V_{i,j}\\
V_{ic}  &=& \sum_{ {{i \in C_k} \atop {j \in C_l}} \atop k \neq l} V_{i,j}
\end{eqnarray}

\begin{figure}
\centerline{\epsfig{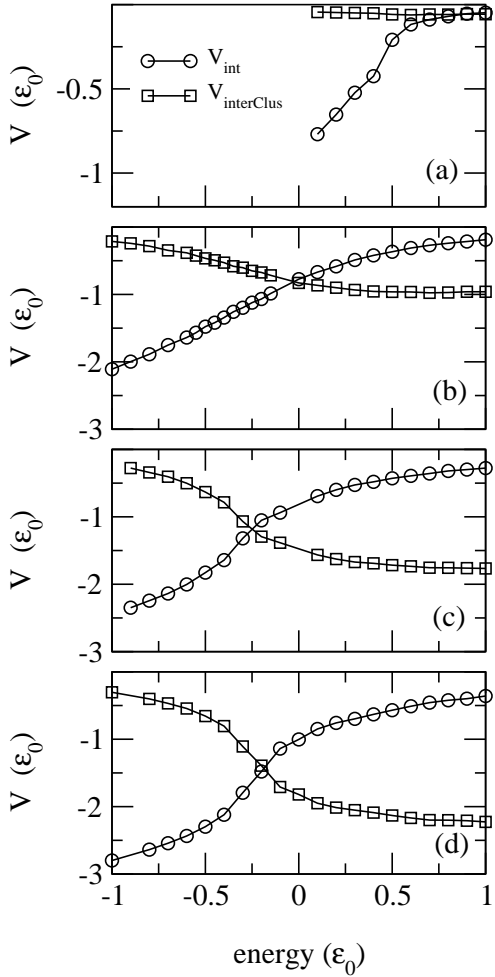}}
\caption{$V_{int}$ (empty circles) and $V_{ic}$ (empty squares) as a function
of the system total energy are shown. Densities: $\rho= 0.01, 0.20, 0.50$, and $0.80
\sigma^{-3}$ are displayed in panels (a), (b), (c), and (d), respectively}\label{figKVC}
\end{figure}

$V_{int}$ stands for the mean potential felt by a particle in a cluster
due to its interaction with the other members of the same cluster.
On the other hand, $V_{ic}$ represents the mean interaction
potential felt by a particle due to its interactions with particles
belonging to the other clusters.
This magnitudes are displayed in Fig.\ref{figKVC} for densities
$\rho=0.01, 0.20, 0.50$, and $0.80 \sigma^{-3}$ in panels (a) to (d)
respectively.

The general trend displayed by $V_{int}$ and $V_{ic}$  
is easy to understand.
At any given density, for low energies, a big ECRA cluster can be found
having a large binding energy. In addition, as no many other clusters beside 
the biggest one exist, $V_{ic} \sim 0$.
At high energies, the ECRA
partitions turn out to present a rather high multiplicity of light clusters.
Consequently $V_{int} \rightarrow 0$ in this limit and $V_{ic}$ absolute
value increases.

After a close inspection of Fig.\ref{figKVC},
one can  realize that, for any given density, the energy
at which $V_{int} = V_{ic}$, happens to be $E_*(\rho)$, i.e. the energy
at which power-law like mass distributions appear. 
This means that scale-free ECRA mass distributions have the following
property: a balance is established between the mean potential energy a
particle, that belongs to a given cluster, feels due to its interaction with other
members of the same cluster
and the one associated with its interaction with the rest of the particles
in the system. 
This potential energy balance, that does not allow to distinguish 
contributions from inside or outside of a cluster,
is reminiscent of the vanishment of the chemical potential and
surface tension terms that takes place at the critical point within the
Fisher droplet model framework \cite{fisher}.

\begin{figure}
\centerline{\epsfig{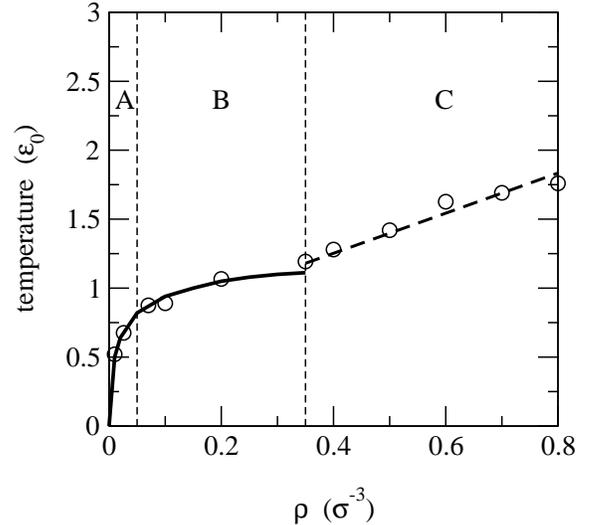}}
\caption{Phase diagram for a 147 particle Lennard Jones drop.
Cluster-based and $c_v$-based determination of the coexistence line is
marked with empty circles, and full line respectively.
The dashed line is included to guide the eye.}\label{figPD}
\end{figure}

Gathering all the information obtained so far, the phase diagram 
for our 147-particle Lennard-Jones drop can be built.
 In Fig.\ref{figPD} the resulting phase diagram is
presented. The empty circles are $(T_*,\rho)$ points obtained from the
aforementioned cluster analysis. The full line is an estimation of the
coexistence line, as obtained from the analysis of the system specific 
heat, $c_v$ (see Eq.\ref{eqFluc}). In this case, for densities for which the respective
CC displays a loop, $T(E_*)$ has been taken as the average temperature between the two values
corresponding to the location of the $c_v$ singularities. For denser cases, the location of
the maximum of the $c_v$ has been taken.  No reliable estimation of the energy
 that maximizes $c_v$ could be achieved for $\rho > 0.35 \sigma^{-3}$.

From the figure, it can be seen that the symmetry between low and high densities,
 characteristic for the infinite size limit,
is lost in the finite size system. An almost linear raising branch, for
$\rho > \rho^c$, can be seen instead. For every density value, below the
depicted coexistence line, a liquid-like ECRA structure can be identified,
whereas a vapor behavior can be observed above it. 

It is interesting to notice that three density regions can be
identified. The region labelled $A$ ($\rho\le0.05 \sigma^{-3}$) presents the signals
expected for a first order phase transition occurring in a finite system:
the corresponding caloric curve shows a loop that can be associated to 
a negative specific heat, and a structural transition can be recognized
(Fig.\ref{figRij}(a)). These region is the only one in which the
available volume is large enough for the system to fragment into a set
of non-overlapping drops.

On the other hand, in region $C$ ($\rho > 0.35 \sigma^{-3}$), where the
system is rather compressed, and MST algorithm recognizes just one big
fragment, a second
order transition seems to occur in phase space, whenever the transition line is
crossed. No anomaly in the respective caloric curve is observed, no qualitative 
changes in the configuration statistical properties is reported (Fig.\ref{figRij}(b)),
and the calculated critical exponents, $\tau$, and $\gamma$, are in good agreement 
with 3D-Ising (liquid-gas) universality class.

In between those two regions, region $B$ 
($0.05\sigma^{-3} < \rho < 0.35\sigma^{-3}$) can be identified.
The finite size of the system plays a major role in this
density range. Even a sensible $\tau$ value suggests that physically 
meaningfull scaling properties are present in the system, the corresponding 
$\gamma$ exponent values are too low to classify the transition as a continuous one.

\section{Conclusions}\label{secConc}

In this paper we have undertaken the analysis of thermodynamical
properties of a finite classical system confined in a volume.
Apart from the basic interest on such a problem, it might be 
relevant on the frame of the description of fragmenting system according
to statistical models.

We have been able to find, that from a coordinate space stand point the
equation of state of such a system is quite simple. There is a maximum
value of the density ($\rho \sim 0.05 \sigma^{-3}$) up to which the system can
undergo a first order phase transition. In this region, the associated
CC displays a loop and the thermal response function attains negative
values. This feature allowed us to define a transition curve.
For higher densities, there is simply no room enough to allow the system
to develop well defined internal surfaces, i.e. only one big drop can be
detected. 

When one turns into a description of fragments in which correlations in
{\bf p}-space are taken into account, a much reacher structure appears.
Now, even for densities bigger than $\rho \sim 0.05 \sigma^{-3}$  transitions
from u-shaped 
mass spectra to exponentially decaying ones are signaled by the appearance of
scale-free mass distribution of ECRA clusters. However, from the analysis of
 critical exponents, $\tau$ and $\gamma$, related to the displayed
 power-law distribution, a further classification in density ranges can
 be obtained.

In the abovedefined region $A$, the obtained $\tau$ values are too big to be
related to a true critical behavior. For densities 
$0.05 \sigma^{-3} < \rho < 0.35 \sigma^{-3}$, eventhough the values of
$\tau$ exponents are in good agreement with the corresponding 3D-Ising
universality class, the value attained by $\gamma$ critical exponents
came out to be too low.
Finally, for densities above $\rho \sim 0.35 \sigma^{-3}$, both, $\tau$
and $\gamma$, are quite close to the accepted values for the 3D-Ising
universality class.

This is of particular interest for the statistical model approach 
because by definition, at freeze out, an ensemble of well defined
fragments is to be constructed. Such assumption implicitly locates the
system under study in region $A$, where first order phase transition is
to be expected. It is then clear that in the frame of such approaches
regions $B$ and $C$ are excluded from the analysis.

A natural continuation of this work is to analyze the relation 
between fragmentation inside the container volume and the corresponding
asymptotic mass spectra when walls are removed. This work is currently
under progress.

Finally, we would like to mention that the behavior of the equation of
state for small systems has been undertaken in Ref.~\cite{campi1y2}. The
results obtained in that work differs from ours (critical exponents) but they 
were based on properties of MSTE
fragments, which we have disregarded favoring the ECRA approach because
of the reasons stated right after Fig.\ref{figClusters}.


\begin{thebibliography}{99}
\bibitem{aclara}
It should be noticed that, when dealing with unconstrained expanding
systems, the quantities appearing in Eq. \ref{eqAclara} should be
calculated at the time of fragment formation (see~\cite{Strachan97}).
\end{thebibliography}
\end{document}